\documentclass[aps,prb,nobibnotes,reprint,twocolumn,amsmath,amsfonts,amssymb]{revtex4-1}
\usepackage{graphicx}
\usepackage{caption}
\usepackage{bm} 
\usepackage{color}
\usepackage{cancel}
\usepackage{hyperref}

\begin{document}


\title{Construction of linearly independent non-orthogonal AGP states}

\author{Rishab Dutta}
\affiliation{Department of Chemistry, Rice University, Houston, TX 77005}

\author{Guo P. Chen}
\affiliation{Department of Chemistry, Rice University, Houston, TX 77005}

\author{Thomas M. Henderson}
\affiliation{Department of Chemistry, Rice University, Houston, TX 77005}
\affiliation{Department of Physics and Astronomy, Rice University, Houston, TX 77005}

\author{Gustavo E. Scuseria}
\email[Correspondence email address:]{guscus@rice.edu}
\affiliation{Department of Chemistry, Rice University, Houston, TX 77005}
\affiliation{Department of Physics and Astronomy, Rice University, Houston, TX 77005}


\begin{abstract}
We show how to construct a linearly independent set of antisymmetrized geminal power (AGP) states, which allows us to rewrite our recently introduced geminal replacement models as linear combinations of non-orthogonal AGPs. 
This greatly simplifies the evaluation of matrix elements and permits us to introduce an AGP-based selective configuration interaction method, which can reach arbitrary excitation levels relative to a reference AGP, balancing accuracy and cost as we see fit.
\end{abstract}

\maketitle

\section{Introduction} \label{sec: intro}

Most wave function methods require expanding the wave function in a many-body basis. The basis of a many-electron wave function is typically chosen as a set of orthonormal Slater determinants, constructed as particle-hole excitations out of a reference determinant. This approach has the advantage that the Hamiltonian matrix is sparse, matrix elements are easy to compute, and for weakly-correlated systems, the expansion coefficients can be factorized using, e.g., some variant of coupled cluster theory.\cite{BartlettBook} 
For strongly-correlated systems, however, Slater determinants and, in turn, molecular orbitals are not the most efficient basis and building blocks, respectively. While we can, in principle, use any basis we wish, we often resort to Slater determinants even when they may not be optimal, largely because there are not many alternatives that facilitate easy computation.

In a series of papers,\cite{TomAGPCI2019,TomAGPRPA2020,
GRAGP2020,AGPQC2021,ExpAGP2020} our group has explored abandoning Slater determinants as the many-electron basis and proposed instead working with wave functions where the basic building blocks are two-electron functions called 
geminals, \cite{Coleman1963} instead of one-electron spin-orbitals. Specifically, the exact wave function can be written in the basis of identical geminal product states known as the antisymmetrized geminal power (AGP), \cite{Coleman1965} which can be obtained as the number-projected Bardeen-Cooper-Schrieffer (PBCS) wave function.\cite{RingBook}
AGP is variationally more flexible than a Slater determinant and is the simplest wave function supporting off-diagonal long-range order in a number-conserving framework. \cite{Yang1962}

The question then becomes how one is to correlate AGP. One possibility is what we call a symmetry \textit{break-correlate-project} approach\cite{Duguet2015,Ten-no2016,Jacob2017,Ethan2017,
Ethan2018,Ethan2019} in which we first correlate the symmetry-broken BCS mean-field, then project the resulting correlated wave function.  Alternatively, we can imagine a symmetry \textit{break-project-correlate} technique in which we directly correlate an AGP state, essentially by expanding the wave function in terms of AGP and states replacing one or more geminals.\cite{GRAGP2020} 
While this construction works well and accurately describes strong pairing interactions, it becomes increasingly challenging as we replace more geminals.

In this paper, we explore an alternative perspective. Rather than writing the wave function in terms of AGP and operators that replace geminals, we write the wave function as a linear combination of AGPs (LC-AGP), \cite{Uemura2015,Uemura2019,GRAGP2020} which has the advantage that matrix elements between two different AGP states are simple. 
LC-AGP is related to the symmetric tensor decomposition of the exact wave function, \cite{Uemura2015,GRAGP2020} and non-orthogonal construction of AGP bases is natural due to the close connection between AGPs and elementary symmetric polynomials (ESPs).\cite{Khamoshi2019}

Given a set of AGPs, optimizing the expansion coefficients in LC-AGP is easily done as a non-orthogonal configuration interaction. \cite{NOCIUrban1969} 
On the other hand, optimizing the geminals of the AGPs is quite difficult; the problem becomes somewhat akin to that of a non-orthogonal multi-configurational self-consistent field.\cite{ESTBook} 
Numerical issues with full variational optimization of LC-AGP, which would require optimizing both configuration and geminal coefficients of individual states, have been observed in the literature.\cite{Uemura2015,GRAGP2020} 

For this reason, we pursue a different goal in this paper. Rather than optimizing AGP states in an LC-AGP, we wish to construct a linearly independent set of non-orthogonal AGP states such that the LC-AGP identically reproduces the geminal replacement theories we have considered in previous work. 
This would allow us to develop a selective configuration interaction (SCI)\cite{SCIHackmeyer1969,SCIBuenker1975} algorithm designed for a non-orthogonal manifold to variationally approximate different geminal replacement methods using as few AGPs as possible. To the best of our knowledge, all of the SCI schemes in the literature are concerned with orthonormal Slater determinants, except one in which orthonormal cluster states were employed as the many-electron basis. \cite{SCIMayhall2020}

This paper is organized as follows. In section~\ref{sec: lcagp}, we present the LC-AGP wave function. Section~\ref{sec: gen} discusses construction of the aforementioned linearly independent non-orthogonal set of AGPs. 
In section~\ref{sec: sci}, we show how to remove energetically less important states before concluding in section~\ref{sec: last}.


\section{LC-AGP} \label{sec: lcagp}

An AGP state with $n$ electron pairs is defined as
\begin{equation} \label{eq: agp}
|n; \: \mu \rangle
= \frac{1}{n!} \left( \Gamma_\mu^\dagger \right)^n |-\rangle,
\end{equation}
where $\mu$ labels this AGP amongst a manifold of AGPs, $|-\rangle$ represents the physical vacuum state, and 
$\Gamma_{\mu}^\dagger$ is a geminal creation operator, which in its canonical basis can be expanded as
\begin{equation} \label{eq: gem}
\Gamma_{\mu}^\dagger 
= \sum_{p=1}^{m} \: \eta_p^{\mu} \: P_p^\dagger,
\end{equation}
where $m$ is the number of spatial orbitals, $\eta_p^{\mu}$ is the $p$-th geminal coefficient of this AGP, and 
\begin{equation} \label{eq: pdag}
P_p^\dagger 
= c_p^\dagger \: c_{\bar{p}}^\dagger
\end{equation}
is the pair creation operator.
Here, $c_p^\dagger$ is a fermion creation operator and spin-orbitals $\bar{p}$ and $p$ are \textit{paired}, i.e., they are defined by conjugate pairs in the canonical unitary congruence transformation of an antisymmetric matrix.
\cite{Hua1944,TomAGPRPA2020}
For notational simplicity, we will use $|n; \:\mu \rangle$ and 
$|\mu \rangle$ interchangeably to denote an AGP. The state $|n \rangle$ without additional indices should be understood as the reference AGP state.

The LC-AGP wave function is
\begin{equation} \label{eq: lcagp}
| \Psi \rangle 
= \sum_{\mu=1}^{R} \: C_{\mu} \: | \mu \rangle,
\end{equation}
where $R$ is the number of AGPs and an $n$-pair AGP $|\mu \rangle$ is created by the geminal creation operator defined in eq.~\eqref{eq: gem}.
Given a set of AGPs, the LC-AGP coefficients $C_\mu$ can be solved as a generalized eigenvalue problem
\begin{equation} \label{eq: gep}
\mathbf{H \: C = M \: C \: E},
\end{equation}
where
\begin{subequations} \label{eq: mat}
\begin{align}
H_{\mu \nu}
&= \langle \mu | \: H \: | \nu \rangle,
\\
M_{\mu \nu}
&= \langle \mu | \nu \rangle,
\end{align}
\end{subequations}
and \textbf{E} is a diagonal matrix containing the ground and excited
state energies, and the columns of \textbf{C} are the corresponding
LC-AGP coefficients.

The AGP wave function of eq.~\eqref{eq: agp} has seniority symmetry, \cite{RingBook} which means that all electrons are paired. Previous work has shown that seniority can be a useful tool for organizing Hilbert space in the presence of strong correlation. \cite{Seniority2011,SeniorityCC2014,Seniority2015}
Doubly-occupied configuration interaction (DOCI), \cite{DOCI1967,MCSCFClementi1967,GMOHall1997,
Kollmar2003} the most general zero-seniority wave function, can account for a large fraction of the correlation energy in many strongly-correlated systems. \cite{Seniority2011,AyersJCTC2013,Stein2014,
SeniorityCC2014,BultinckDOCI2015}
In this work, we will be attempting to reach DOCI accuracy with AGP-based methods. However, DOCI has combinatorial cost, so benchmark DOCI results are computationally demanding.  

We therefore limit ourselves to the pairing or reduced BCS Hamiltonian,
\begin{equation} \label{eq: ham}
H = \sum_p \epsilon_p \: N_p - G \: \sum_{pq} \: P_p^\dagger P_q,
\end{equation}
where indices $p$ and $q$ label spatial orbitals or \textit{levels}, the one-body interaction is assumed to be $\epsilon_p = p$, and the two-body interaction may be repulsive ($G < 0$) or attractive ($G > 0$); $P_q$ is a pair annihilation operator and is the adjoint of the pair creation operator $P_q^\dagger$ of eq.~\eqref{eq: pdag}, and the number operator $N_p$ is defined as
\begin{equation}
N_p = c_p^\dagger \: c_p + c_{\bar{p}}^\dagger \: c_{\bar{p}}.
\end{equation}
The pair and number operators form a representation of generators of the 
$su(2)$ algebra
\begin{subequations}
\begin{align}
\label{eq: com1}
[P_p^\dagger, P_q]
&= \delta_{pq} \: (N_p - 1), \\
\label{eq: com2}
[N_p, P_q^\dagger] 
&= 2 \: \delta_{pq} \: P_q^\dagger.
\end{align}
\end{subequations}
Although simplistic, this Hamiltonian shows non-trivial physics in the attractive regime,\cite{BCSCC2014} which traditional quantum chemistry methods fail to describe.\cite{pECCD2015,PoST2016,BCSCC2014} 
In contrast, AGP and AGP-based methods are able to capture most of the correlation energies systematically. \cite{TomAGPCI2019,TomAGPRPA2020,GRAGP2020,
AGPQC2021,ExpAGP2020}

Because the reduced BCS Hamiltonian has seniority as a symmetry, it is exactly solved by DOCI. However, the Hamiltonian is exactly solvable through a set of non-linear Richardson-Gaudin equations, \cite{RG2004,Richardson1963,Richardson1964}
which give access to exact energies even when the number of levels is large.  
The eigenstates of this Hamiltonian, also known as Richardson-Gaudin states, have been used to approximately solve the molecular Hamiltonian.\cite{RG2020,RGRDM2020,RGSD2020,RGTDM2020}

As we have noted, LC-AGP requires us to compute overlaps between AGPs to build the metric \textbf{M}, and we need transition reduced density matrices (RDMs) between different AGPs to compute the Hamiltonian matrix \textbf{H}. For the reduced BCS Hamiltonian of eq.~\eqref{eq: ham}, we need
\begin{subequations} \label{eq: tdm}
\begin{align}
Z_{\mu \nu, p}^{1, 1}
&= \langle \mu | \: N_p \: | \nu \rangle,
\\
Z_{\mu \nu, pq}^{0, 2}
&= \langle \mu | \: P_p^\dagger P_q \: | \nu \rangle,
\end{align}
\end{subequations}
so that the Hamiltonian matrix is
\begin{equation} \label{eq: ham2}
\langle \mu | \: H \: | \nu \rangle
= \sum_p \epsilon_p \: Z_{\mu \nu, p}^{1, 1}
- G \: \sum_{pq} \: Z_{\mu \nu, pq}^{0, 2}.
\end{equation}

The AGP overlaps and transition RDMs can be computed as ESPs.\cite{Khamoshi2019}
The computational complexity for $\mathbf{M}$ and $\mathbf{H}$ are $\mathcal{O}(m R^2)$ and $\mathcal{O}(m^4 R^2)$, respectively. In the regime where all 
$\bm{\eta}^\mu$ coefficients are different, the latter can be reduced to $\mathcal{O}(m^3 R^2)$ via the \textit{reconstruction formulae},
in which higher-order RDMs are expressed as linear combinations of lower-order ones. \cite{Khamoshi2019} 
Relevant expressions are presented in Appendix~\ref{app: rf}.
Alternatively, we may express the AGPs as PBCS states and evaluate the AGP transition RDMs as BCS transition RDMs integrated over a gauge angle; see Appendix~\ref{app: pbcsrdm}.
The costs of computing $\mathbf{M}$ and $\mathbf{H}$ scale as $\mathcal{O}(lmR^2)$ and $\mathcal{O}(lm^2R^2)$, respectively, where $l$ denotes the size of the numerical quadrature for the gauge integration.
In fact, we can evaluate $\mathbf{Z}^{11}$ in $\mathcal{O} (l m R^2)$ and then \textbf{H} in $\mathcal{O} (m^2 R^2)$ by combining the ideas discussed above.

The most general seniority-zero geminal product wave function, the antisymmetrized product of interacting geminals (APIG)\cite{APIG1971,AyersCTC2013,AyersJCTC2013}
\begin{equation} \label{eq: apig}
|\mbox{APIG} \rangle
= \prod_{\mu = 1}^{n} \: \Gamma_{\mu}^\dagger \: |-\rangle
\end{equation}
is a promising wave function for strong correlation. In general, APIGs are computationally complex to work with since the expansion coefficients of an APIG in a Slater determinant basis are permanents. \cite{AyersCTC2013,AyersJCTC2013,ThesisJohnson2014}
However, an arbitrary APIG for $n$ pairs can be expressed as a sum of $2^{n-1}$ non-orthogonal AGPs, as shown in the Appendix of Ref.~\onlinecite{GRAGP2020}.
Incidentally, the exact ground state of the reduced BCS Hamiltonian is a special form of APIG,\cite{AyersCTC2013,TomAGPRPA2020} which means $2^{n-1}$ non-orthogonal AGPs are in principle enough to find the exact ground state of the reduced BCS Hamiltonian. In general, these AGPs have complex-valued geminal coefficients, and their optimization is exceptionally cumbersome.
In this work, we prefer to avoid these difficulties by working with a fixed basis of AGPs, the construction of which we will now describe.


\section{Generating AGP states} \label{sec: gen}

Here we show how to construct a linearly independent non-orthogonal set of AGPs from any reference AGP. We will first discuss the construction and then discuss how to reproduce different geminal replacement models with LC-AGP. 

\subsection{Cosenior transformation of AGP} \label{subsec: agptrans}

As shown in Ref.~\onlinecite{ExpAGP2020}, one AGP can be transformed to another with modified geminal coefficients via  
\begin{equation} \label{eq: ej1}
|n ; \: \mu \rangle
= e^{J_1^\mu} \: |n\rangle,
\end{equation}
where
\begin{equation}
J_1^\mu = \sum_p \: s_p^\mu \: N_p,
\end{equation}
with the corresponding geminal coefficient transformed as
\begin{equation}
\eta^\mu_p 
= e^{2 s^\mu_p} \: \eta_p
= \alpha^\mu_p \: \eta_p.
\end{equation}
Conversely, any AGP $|n ; \: \mu \rangle$ can be generated from a reference AGP 
$|n \rangle$ through eq.~\eqref{eq: ej1}, provided that $s_p^\mu$ is 
complex-valued and $|n ; \: \mu \rangle$ and $|n \rangle$ are \textit{cosenior}, i.e., they share the same canonical orbitals or natural orbitals of the geminal. 
Eq.~\eqref{eq: ej1} for cosenior AGPs can be considered as an analogue of the Thouless theorem for Slater determinants.\cite{ThoulessTheorem}

By expanding the exponential of $J_1^\mu$ in eq.~\eqref{eq: ej1}, it is straightforward to show that the new AGP can be generated by acting a product of shifted number operators on the reference AGP
\begin{equation} \label{eq: prodnbar}
|n ; \: \mu \rangle 
= \xi \prod_p \left(N_p + \beta^\mu_p\right) \: |n\rangle,
\end{equation}
where
\begin{subequations} \label{eq: p1}
\begin{align}
\beta^\mu_p 
&= \frac{2}{\alpha^\mu_p - 1},
\\
\xi 
&= \prod_p \frac{1}{\beta^\mu_p}.
\end{align}
\end{subequations}
We have used the fact that different number operators commute and 
\begin{equation} \label{eq: idem}
N_p^2 = 2 N_p
\end{equation}
for seniority-zero states.
When $s^\mu_p = 0$ for a given $p$, $\alpha_p^\mu = 1$ and $\beta_p^\mu$ is not well defined, but eq.~\eqref{eq: prodnbar} still holds after removing the $p$-th factor from the product sequence. We refer to the remaining $\alpha_p^\mu$ and 
$\beta_p^\mu$ as \textit{pivots} and \textit{shifts}, respectively.

Here we provide an alternative perspective on the same transformation.
An $m$-level, $n$-pair AGP can be represented as an $m$-variable, $n$-degree 
ESP\cite{ESP2011} of $\eta_p \: P_p^\dagger$ acting on the vacuum 
\begin{subequations} 
\begin{align}
\label{eq: esp1}
|n\rangle
&= \sum_{1 \leq p_1 < \cdots < p_n \leq m} \eta_{p_1} \cdots \eta_{p_n} \: 
P_{p_1}^\dagger \cdots P_{p_n}^\dagger \: |-\rangle \\
\label{eq: esp2}
&= S_n^m \: ( \{ \eta_{p_i} P_{p_i}^\dagger \: | \: 1 \leq i \leq n\} ) \: 
|-\rangle
\end{align}
\end{subequations}
since all the pairing creation operators commute with each other. 
Using the recursion formula of an ESP, \cite{SumESP1974,ESP1996} we can partition any AGP as
\begin{equation} \label{eq: part}
|n\rangle 
= \eta_p \: P_p^\dagger \: |n-1\rangle_{-p} 
+ |n\rangle_{-p}, \quad \forall \: p.
\end{equation}
Here the subscript $-p$ means that the level $p$ is excluded from the AGP expansion in eq.~\eqref{eq: esp1}. Eq.~\eqref{eq: part} and the commutation relation of eq.~\eqref{eq: com2} imply that
\begin{subequations} 
\begin{align}
&(N_p + \beta_p^\mu) \: |n\rangle
\nonumber \\
\label{eq: shift1}
&= (2 + \beta_p^\mu) \: \eta_p \: P_p^\dagger \: |n-1\rangle_{-p} 
+ \beta_p^\mu \: |n\rangle_{-p} \\
\label{eq: shift2}
&= \beta_p^\mu \: \big[ (1 + \frac{2}{\beta_p^\mu}) \: \eta_p \: P_p^\dagger \: 
|n-1\rangle_{-p} + |n\rangle_{-p}	
\big].
\end{align}
\end{subequations}
Thus, shifting $N_p$ by $\beta_p^\mu$ pivots the corresponding geminal coefficient $\eta_p$ by 
\begin{equation} \label{eq: pivot}
\alpha_p^\mu 
= 1 + \frac{2}{\beta_p^\mu},
\end{equation}
corroborating eqs.~\eqref{eq: ej1}-\eqref{eq: p1}.

To systematically generate manifolds spanning the DOCI space, we consider AGPs generated by $k$ shifted number operators
\begin{subequations} 
\begin{align}
\label{eq: jbar}
|n; \: p_1 \cdots p_k \rangle
&= \xi' \prod_{i=1}^k \big( N_{p_i} + \beta_{p_i}^\mu \big) \: |n\rangle,
\\
\label{eq: xiprime}
\xi'
&= \prod_{i=1}^k \: \frac{1}{\beta^\mu_{p_i}},
\end{align}
\end{subequations}
where the index $\mu$ should be understood as a composite index of $p_1 \cdots p_k$. 

A few remarks are in order. 
First, when all the shifts are zero, eq.~\eqref{eq: jbar} is not normalizable according to eq.~\eqref{eq: xiprime} and therefore not an AGP;
nevertheless, it becomes an antisymmetrized product of $k$ doubly-occupied orbitals and
geminal power of $(n-k)$ pairs
\begin{equation} \label{eq: sdagp}
\xi'' N_{p_1} \cdots N_{p_k} \: |n\rangle
= \frac{1}{(n-k)!} \: \big( \prod_{i = 1}^{k} \: 
P_{p_i}^\dagger \big)
\: \big( \Gamma^{\dagger} \big)^{n-k} \: |-\rangle
\end{equation}
with the normalization factor
\begin{equation}
\xi'' 
= \prod_{i=1}^k \frac{1}{2 \, \eta_{p_i}}.
\end{equation}
Eq.~\eqref{eq: sdagp} is a basis state of a $J_k$-CI wave function
\begin{equation} \label{eq: jci}
| J_k \mbox{-CI} \rangle 
= \sum_{p_1 < \cdots < p_k} S_{p_1 \cdots p_k} \: N_{p_1} \cdots N_{p_k} 
\: |n\rangle,
\end{equation} 
or equivalently, a $k$-th order geminal replacement ($k$-GR) model
\begin{equation} \label{eq: grci}
| k \mbox{-GR} \rangle 
= \sum_{\mu=1}^{r} \: \lambda_\mu \: \big( \Gamma_{\mu}^{\dagger} \big)^k 
\: | n-k \rangle,
\end{equation}
where the coefficients $S_{p_1 \cdots p_k}$ and $\lambda_\mu$ are related by a symmetric tensor decomposition and $r$ is the symmteric rank of the tensor $S$. \cite{GRAGP2020}
Second, regardless of the choice of the shifts, the number of states generated
by eq.~\eqref{eq: jbar} is ${m \choose k}$, which equals the dimensionality of the $J_k$-CI manifold.
Third, using the theorem in the Appendix of Ref.~\onlinecite{ExpAGP2020},
it can be readily shown that $|n; \: p_1 \cdots p_k \rangle$ is contained in the $J_k$-CI manifold. Consequently, the ${m \choose k}$ states generated by 
eq.~\eqref{eq: jbar} span the $J_k$-CI manifold as long as they are linearly independent. We have indeed observed that a wave function of the form  
\begin{equation} \label{eq: jbarci} 
|\Psi_1 \rangle
= \sum_{p_1 < \cdots < p_k} C_{p_1 \cdots p_k} \: |n; \: p_1 \cdots p_k \rangle
\end{equation}
reproduces $J_k$-CI energies when the set of AGPs are linearly independent. However, linear independence is not always guaranteed, as discussed in Appendix~\ref{app: lin}. Thus, our next task is to modify the construction in such a way as to assure linear independence.

\subsection{The freeze-and-pivot construction} \label{subsec: fanp}


\begin{figure}[t]
\includegraphics[width=0.8\columnwidth]{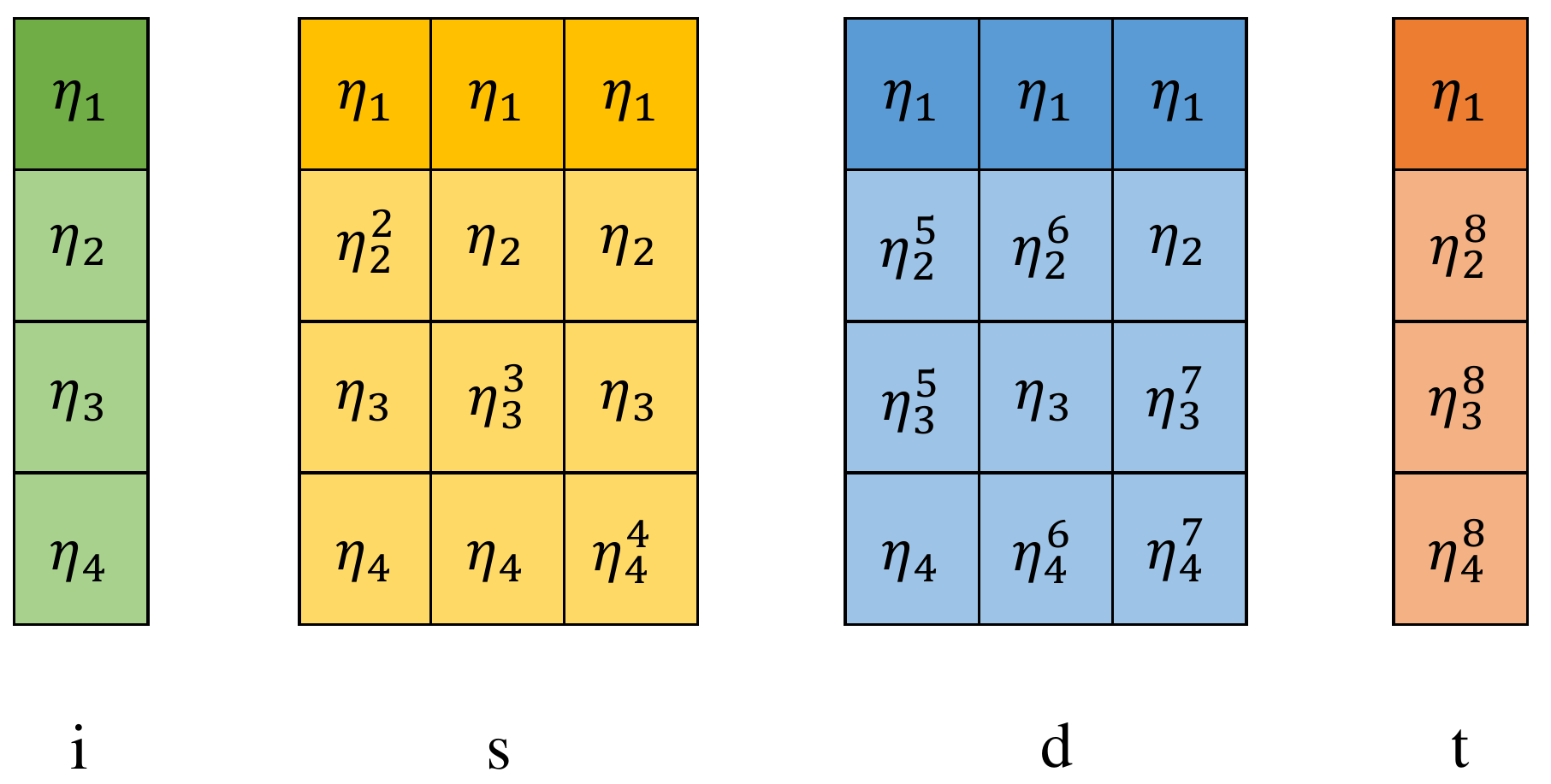}
\caption{
Schematic examples of AGP vectors for different elementary manifolds, for $m=4$. Here the \\ first level is frozen, but freezing any level would be equivalent. The scalars $\eta_p^\mu$ may all be different.
\label{fig: elem}}
\end{figure}


\begin{figure}[b]
\includegraphics[width=\columnwidth]{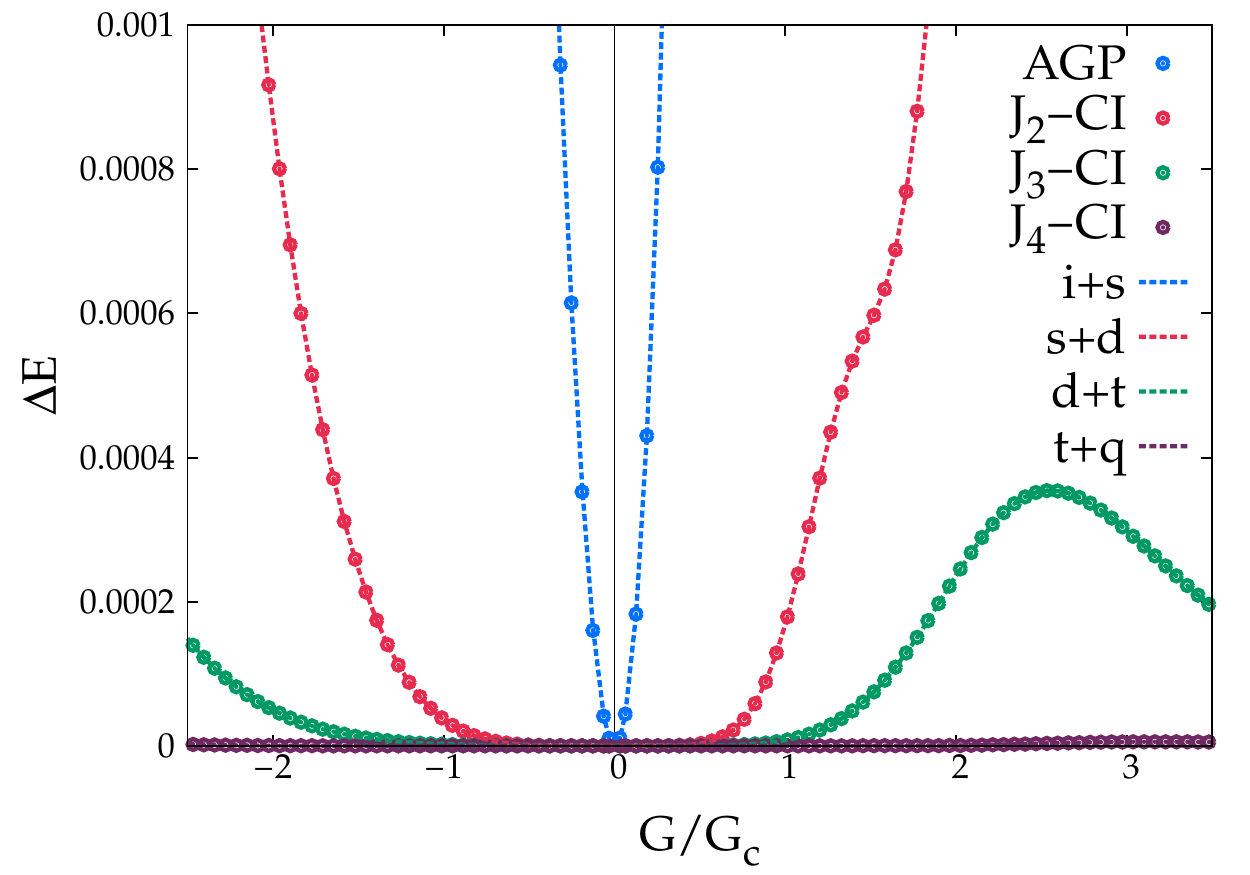}
\caption{
Total energy errors $(E_{method} - E_{exact})$ for different composite manifolds, compared against AGP and various $J_k$-CI methods. 
The system is half-filled 12-level reduced BCS Hamiltonian with critical G value, $G_c \sim 0.3161$. $G/G_C > 1$ is the strong correlation regime for attractive interactions.
\label{fig: en12}}
\end{figure}


\begin{table}[b]
\caption{Percentage of off-diagonal metric elements with absolute values $> 10^{-3}$, for the zero-pivot composite manifolds. The system is the half-filled 20-level reduced BCS Hamiltonian ($G_C \sim 0.2674$). 
\label{tab: met}}
\centering
   \begin{tabular}{lllll}
   \hline
   $G$ & S & D & T & Q \\
   \hline
   -0.60 & 100.00 & 79.52 & 45.40 & 19.82 \\
   -0.30 & 88.95  & 55.03 & 27.40 & 13.69 \\
   0.30  & 100.00 & 81.42 & 52.35 & 27.64 \\
   0.60  & 100.00 & 99.83 & 92.31 & 69.52 \\
   \hline
   \end{tabular}
\end{table}


\begin{figure*}[t]
\includegraphics[width=\columnwidth]{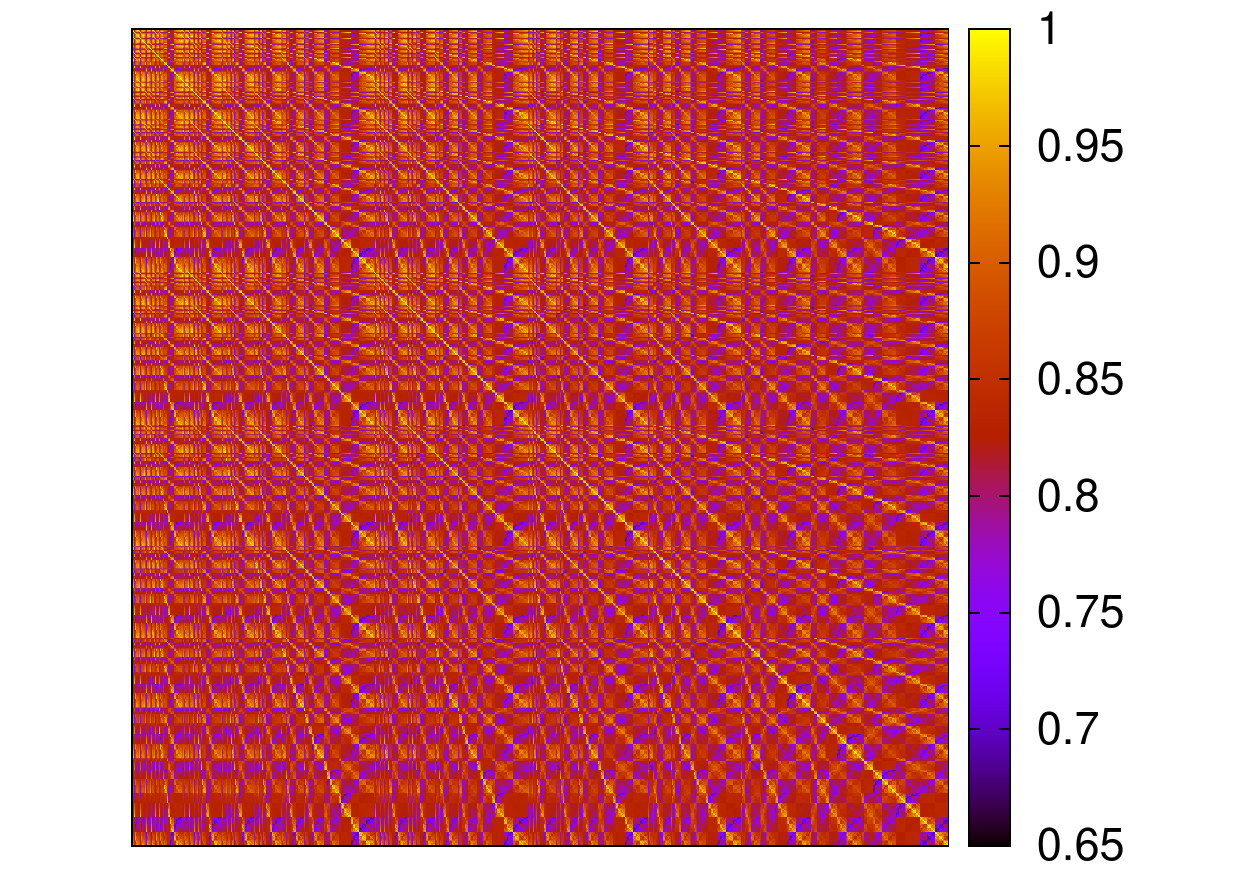}
\hfill
\includegraphics[width=\columnwidth]{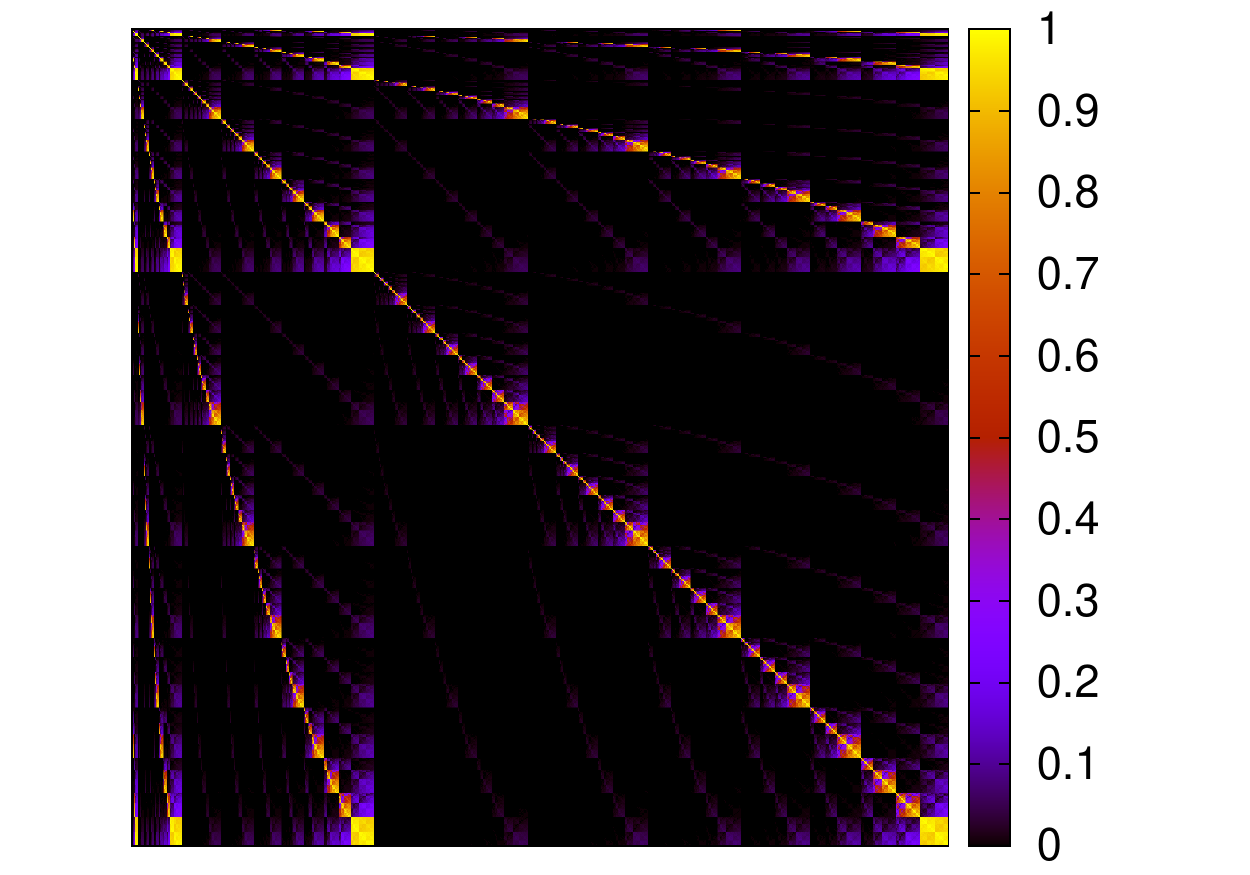}
\caption{Metric matrix heat-maps covering reference AGP to elementary quadruples manifolds (i+s+d+t+q) for the half-filled 16-level reduced BCS Hamiltonian with $G = 0.30$ ($G_C \sim 0.2866$). 
The left and right panels correspond to sign-flip and zero-pivot manifolds, respectively. Only absolute values of the matrix elements are plotted for the sign-flip manifolds. 
For the sign-flip case, only neighboring elementary manifolds may combine to generate linearly independent AGPs whereas combination of any two elementary manifolds generate linearly independent AGPs for the zero-pivot construction. Combinations of more than two elementary manifolds would introduce linear dependence unless some states are removed. 
\label{fig: met}}
\end{figure*}


\begin{figure}[b]
\includegraphics[width=\columnwidth]{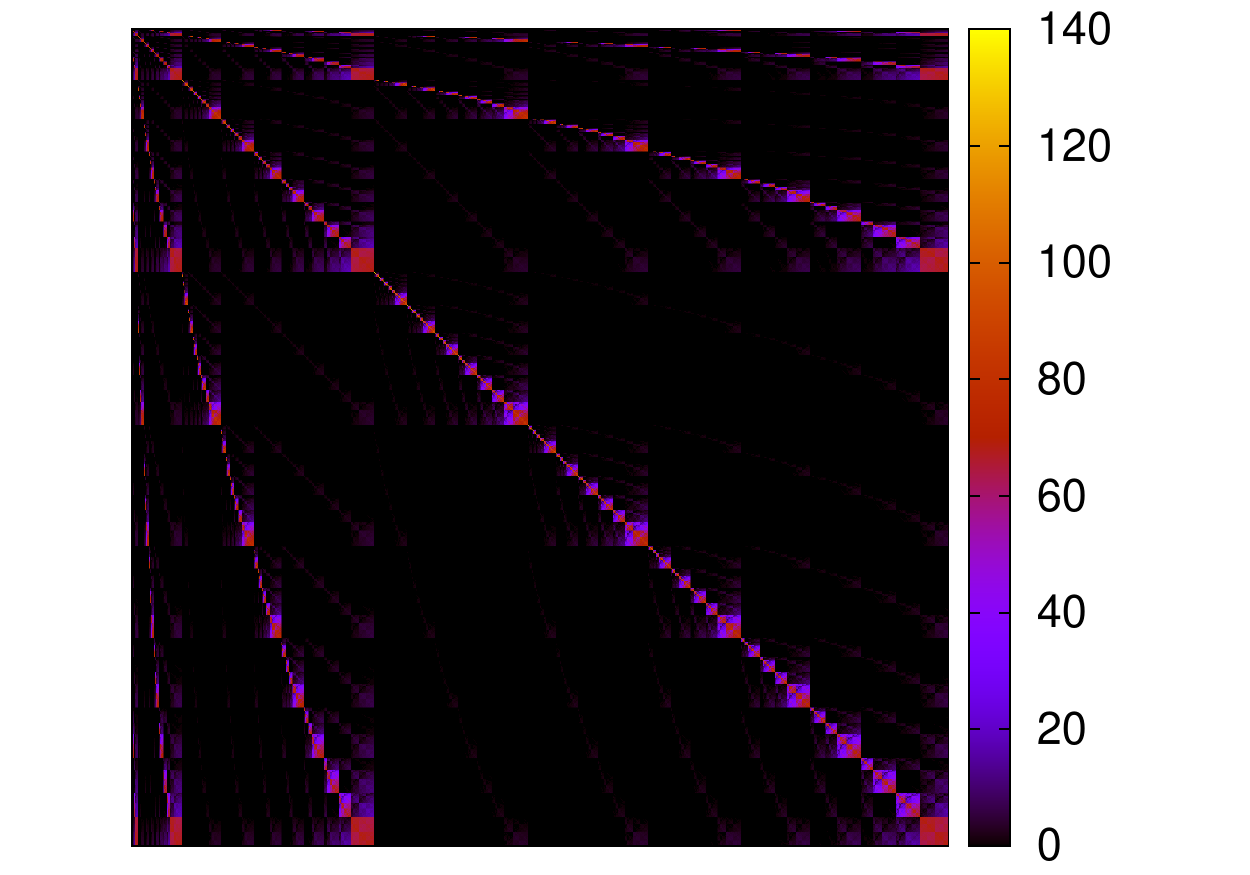}
\caption{Hamiltonian matrix heat-map for the zero-pivot manifolds, covering AGP to elementary quadruples manifolds (i+s+d+t+q). The system is half-filled 16-level reduced BCS Hamiltonian with $G = 0.30$ ($G_C \sim 0.2866$). Only absolute values of the matrix \\ elements are plotted.
\label{fig: ham}}
\end{figure}


We have observed that one way to get a linearly independent set of AGPs is to \textit{freeze} an arbitrary level, excluding it from being pivoted. This is understandable since all the cases of accidental linear dependence examined in
Appendix~\ref{app: lin} are consequences of AGP being an eigenfunction of the total number operator
\begin{subequations} 
\begin{align}
\label{eq: totN}
N &= \sum_{p=1}^m \: N_p,
\\ 
\label{eq: totNonAGP}
N \: |n\rangle 
&= 2n \: |n\rangle.
\end{align}
\end{subequations}
Freezing one level prevents the total number operator from appearing in a linear combination of AGPs of the form given in eq.~\eqref{eq: jbar}. For simplicity, we freeze the first level from now on, and define the \textit{elementary} manifolds using
\begin{itemize}

\item reference AGP (i): $|n\rangle$,

\item elementary singles (s): 
$\{ |n; \: p \rangle \: | \: 1 < p \leq m \}$,

\item elementary doubles (d): 
$\{ |n; \: pq \rangle \: | \: 1 < p < q \leq m \}$,

\end{itemize}
and so on. Figure~\ref{fig: elem} presents an illustration of different elementary manifolds. 

Numerical tests also suggest that for any choice of pivots $\alpha_p^\mu$ (or equivalently, shifts $\beta_p^\mu$), the AGPs in a given elementary manifold are linearly independent of those in a neighboring manifold. We thus define \textit{composite} manifolds using
\begin{itemize}

\item composite singles (S): reference + elementary singles (i + s),

\item composite doubles (D): elementary singles + elementary doubles (s + d),

\item composite triples (T):  elementary doubles + elementary triples (d + t),

\end{itemize}
and so on.
In general, a state in the $k$-th order composite manifold can be written as
\begin{align}
| \Psi_2 \rangle
&= \sum_{1 < p_1 < \cdots < p_{k-1} \leq m} 
C_{p_1\cdots p_{k-1}} \: |n; \: p_1\cdots p_{k-1} \rangle 
\\
&+ \sum_{1 < p_1 < \cdots < p_{k} \leq m} C_{p_1\cdots p_k} \: 
|n; \: p_1 \cdots p_k\rangle.
\end{align}
The dimensionality of this manifold is
\begin{equation} \label{eq: comb}
{m-1 \choose k-1} + {m-1 \choose k} 
= {m \choose k},
\end{equation}
identical to that of $J_k$-CI. We have numerically verified the equivalence between the $k$-th order composite manifold and $J_k$-CI by comparing 
ground state energies after variation.
We show examples of such equivalences in figure~\ref{fig: en12}. 
LC-AGP energies with the composite singles manifold coincide with the reference AGP, because $J_1$-CI does not improve the variationally optimized AGP (although this is not necessarily true when the reference AGP is not optimized). It should also be noted that the highest-order composite manifold ($k=n$) reproduces DOCI or exact energies for the reduced BCS Hamiltonian, even when the reference AGP is not optimized. 

A few special choices of pivots (or equivalently, shifts) are worth some discussion.
When all the pivots are $-1$ (shifts are $-1$), we get the \textit{sign-flip} manifolds, and the AGP vectors, in this case, resemble the power sum decomposition columns of an ESP.\cite{Fischer1994,Lee2016} 
When all the pivots are zero (shifts are $-2$), the AGPs of zero-pivot manifolds amounts to a reduction of levels from the AGP expansion defined in 
eq.~\eqref{eq: esp1}.
The $(-2)$-shift and zero-shift (eq.~\eqref{eq: sdagp}) manifolds represent two limiting cases of 
eq.~\eqref{eq: shift2} since only one term is present in either case. These two shifts are also unique because, in their case, combined states from any two elementary manifolds are observed to form a linearly independent set, not just the neighboring ones. 

The construction described in this section is one of the two main results of this paper. 
We have observed that the procedure outlined above permits us to construct a linearly independent non-orthogonal set of AGPs, even when we work with  different sets of shifts or pivots.
By choosing the appropriate set (e.g., composite doubles) we can exactly reproduce the corresponding geminal replacement models.  The simplicity of evaluating the required matrix elements would permit us to consider fairly complicated theories which replace a large number of geminals, but such theories yield large Hamiltonian and metric matrices. Thus we now turn our attention to pruning the LC-AGP basis in a selective CI (SCI) approach, mentioned in Section~\ref{sec: intro}. We discuss this in Section~\ref{sec: sci}.


\section{SCI with AGP states} \label{sec: sci}


\begin{figure}[b]
\includegraphics[width=0.8\columnwidth]{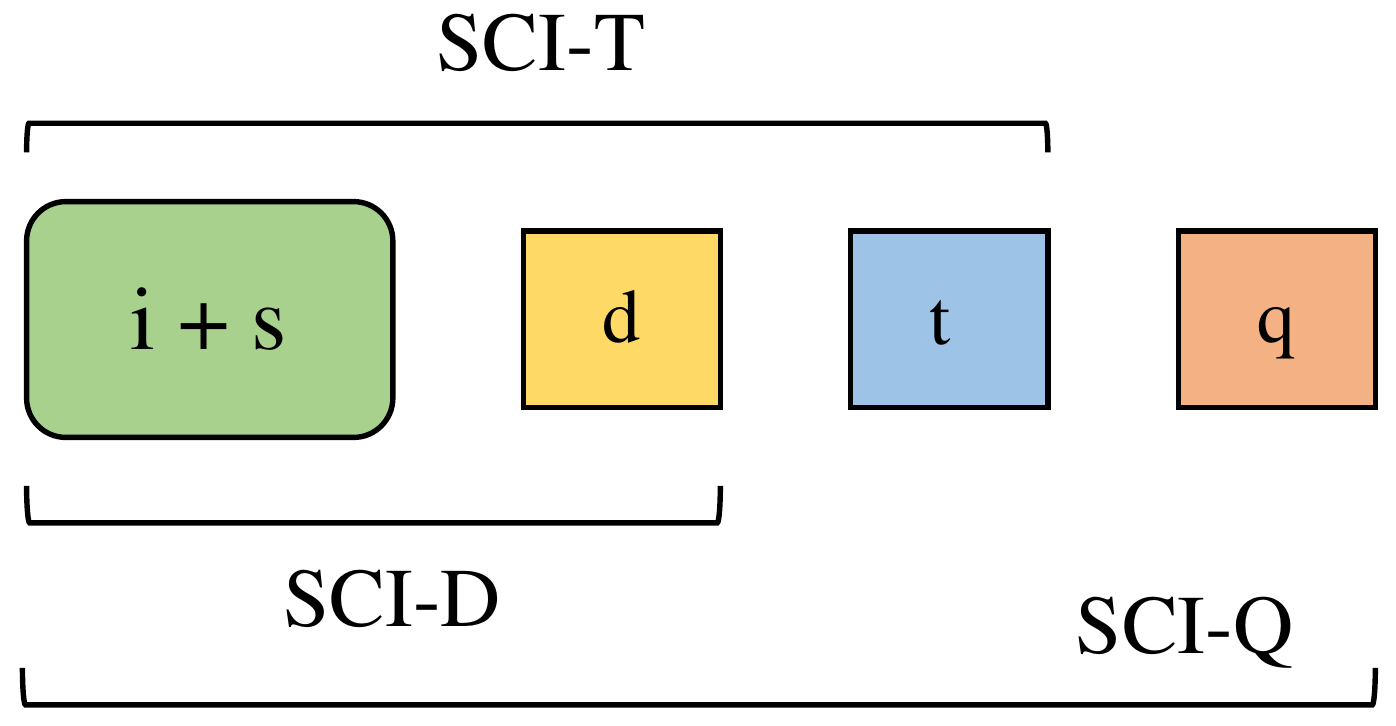}
\caption{ 
Hierarchy of SCI methods based on a common initial model space (i+s) and different candidate spaces.
\label{fig: sci}}
\end{figure}


\begin{figure*}[t]
\includegraphics[width=\columnwidth]{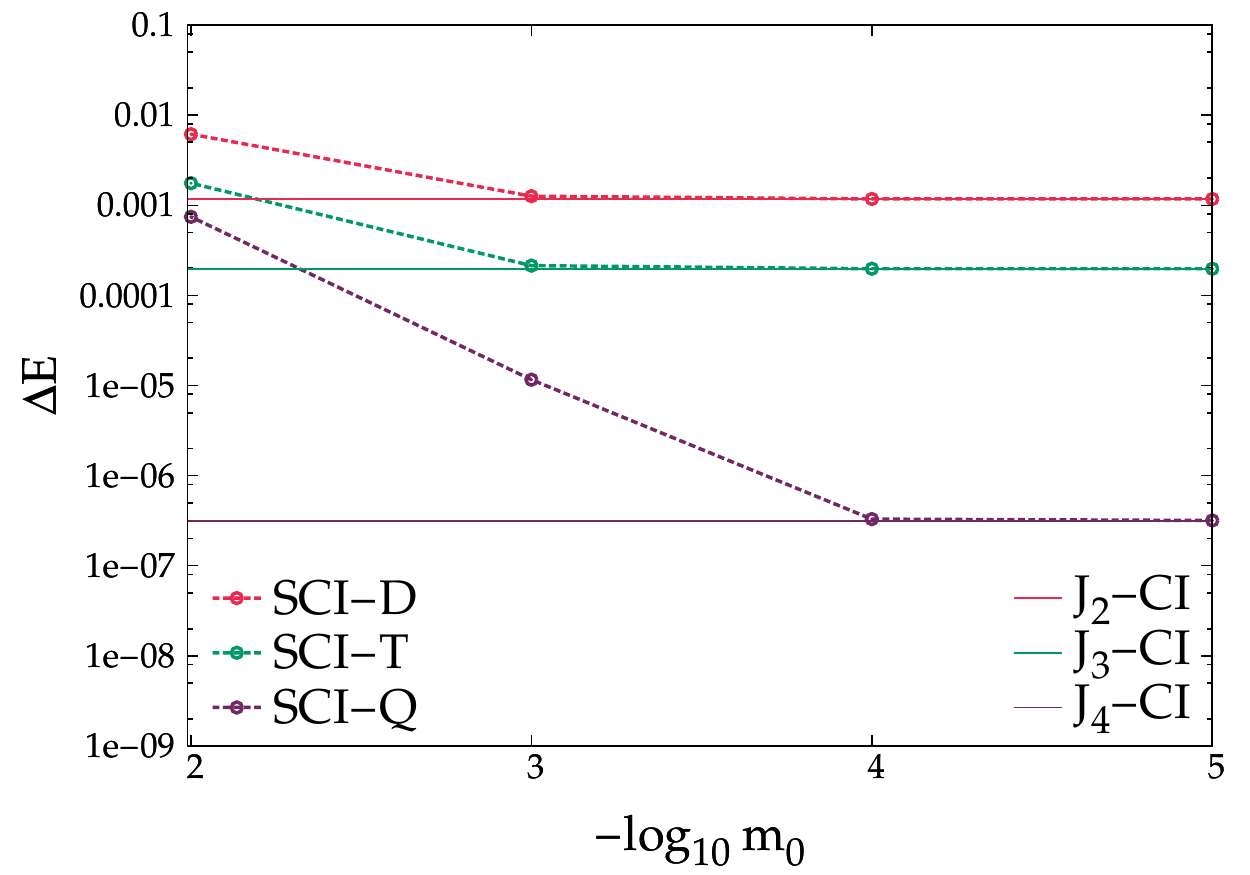}
\hfill
\includegraphics[width=\columnwidth]{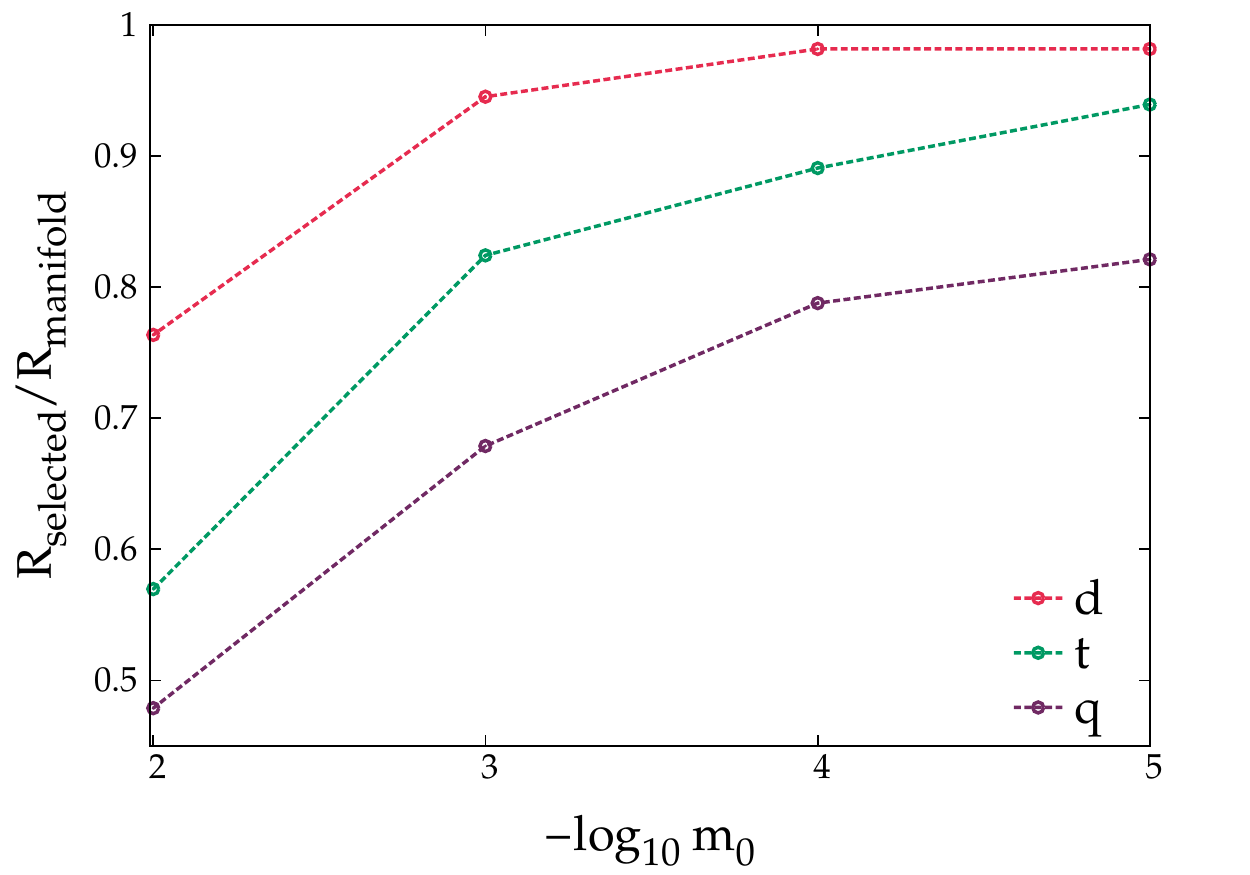}
\caption{
SCI results for the half-filled 12-level reduced BCS Hamiltonian with 
$G = 0.60$ ($G_C \sim 0.3161$). The Hamiltonian threshold is constant at $10^{-12}$ for all the methods. 
The $x$-axes for both the panels are functions of metric thresholds, $m_0$.
In the $y$-axis of the left panel, we plot total energy errors 
$(E_{method} - E_{exact})$ for different SCI methods.
In the $y$-axis of the right panel, we plot the fractions of AGPs selected from different elementary manifolds. For example, the label `d' represents fraction of AGPs selected from the elementary doubles, which is same for all the SCI methods for a given $m_0$.
There are total $55$, $165$, and $330$ states in elementary singles, doubles, and triples, respectively.
\label{fig: en12b}}
\end{figure*}


\begin{figure*}[t]
\includegraphics[width=\columnwidth]{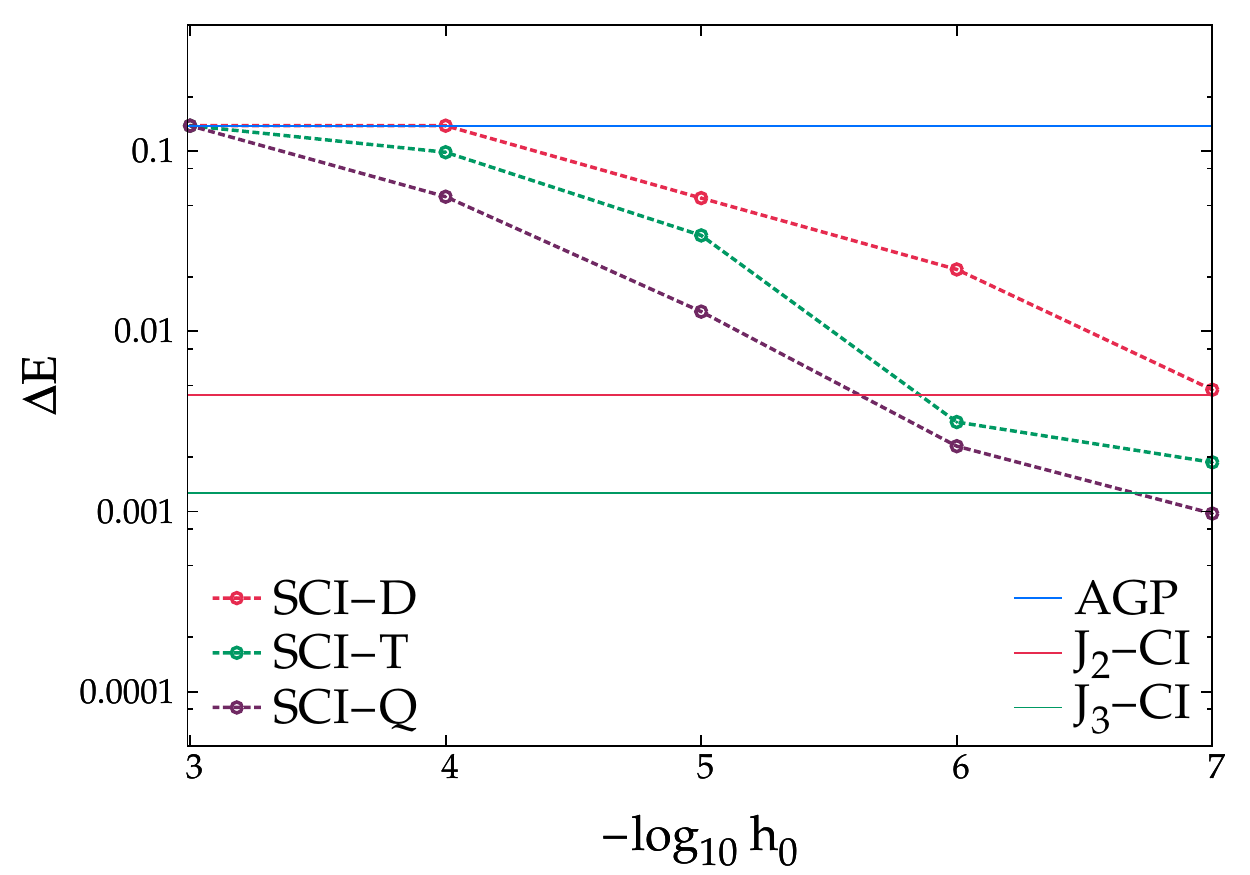}
\hfill
\includegraphics[width=\columnwidth]{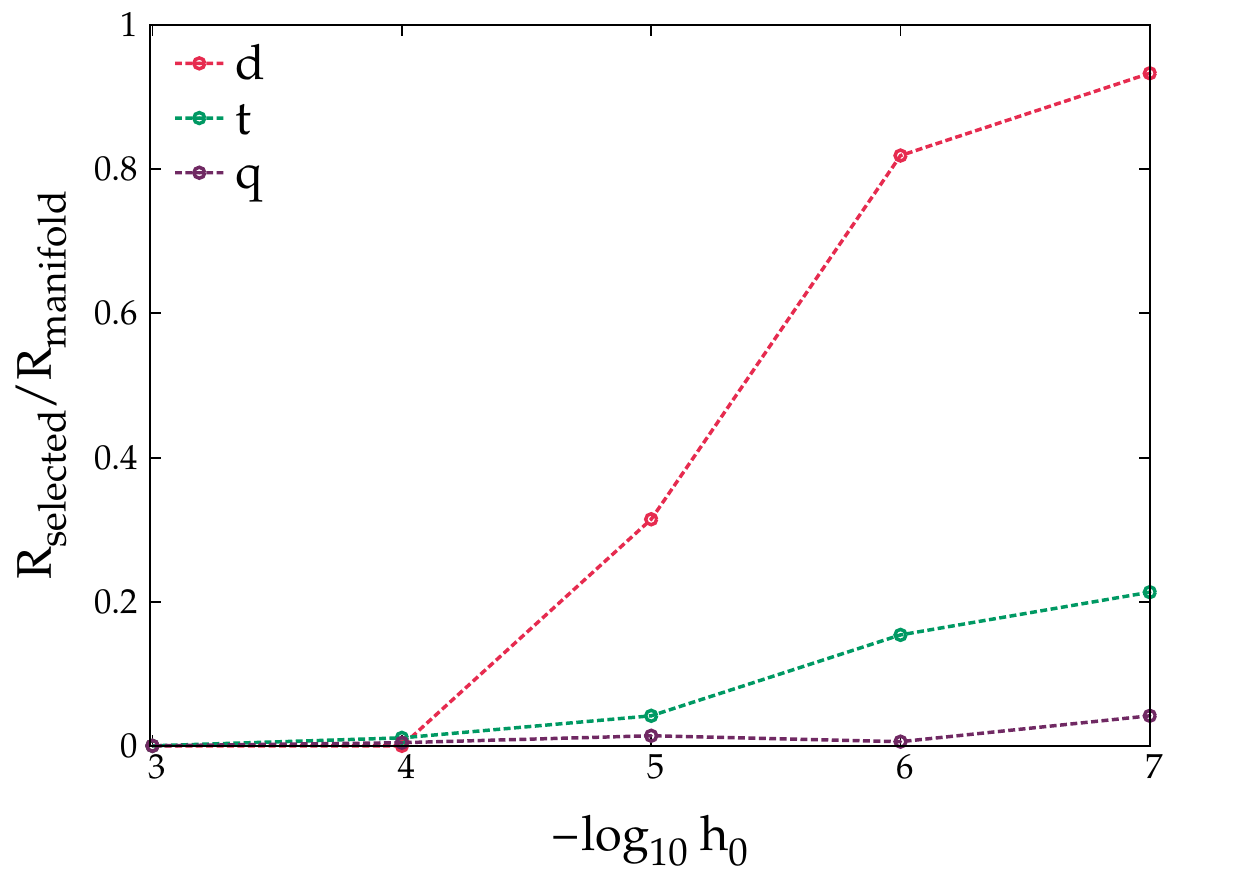}
\caption{
SCI results for the half-filled 16-level reduced BCS Hamiltonian with 
$G = 0.60$ ($G_C \sim 0.2866$). The metric threshold is constant at $10^{-4}$ for all the methods. 
The $x$-axes for both the panels are functions of Hamiltonian thresholds, $h_0$.
In the $y$-axis of the left panel, we plot total energy errors $(E_{method} - E_{exact})$ for different SCI methods.
In the $y$-axis of the right panel, we plot the fractions of AGPs selected from different elementary manifolds. For example, the label `d' represents fraction of AGPs selected from the elementary doubles, which is same for all the SCI methods for a given $h_0$.
There are total $105$, $455$, and $1365$ states in elementary singles, doubles, and triples, respectively.
\label{fig: en16a}}
\end{figure*}


\begin{figure*}[t]
\includegraphics[width=\columnwidth]{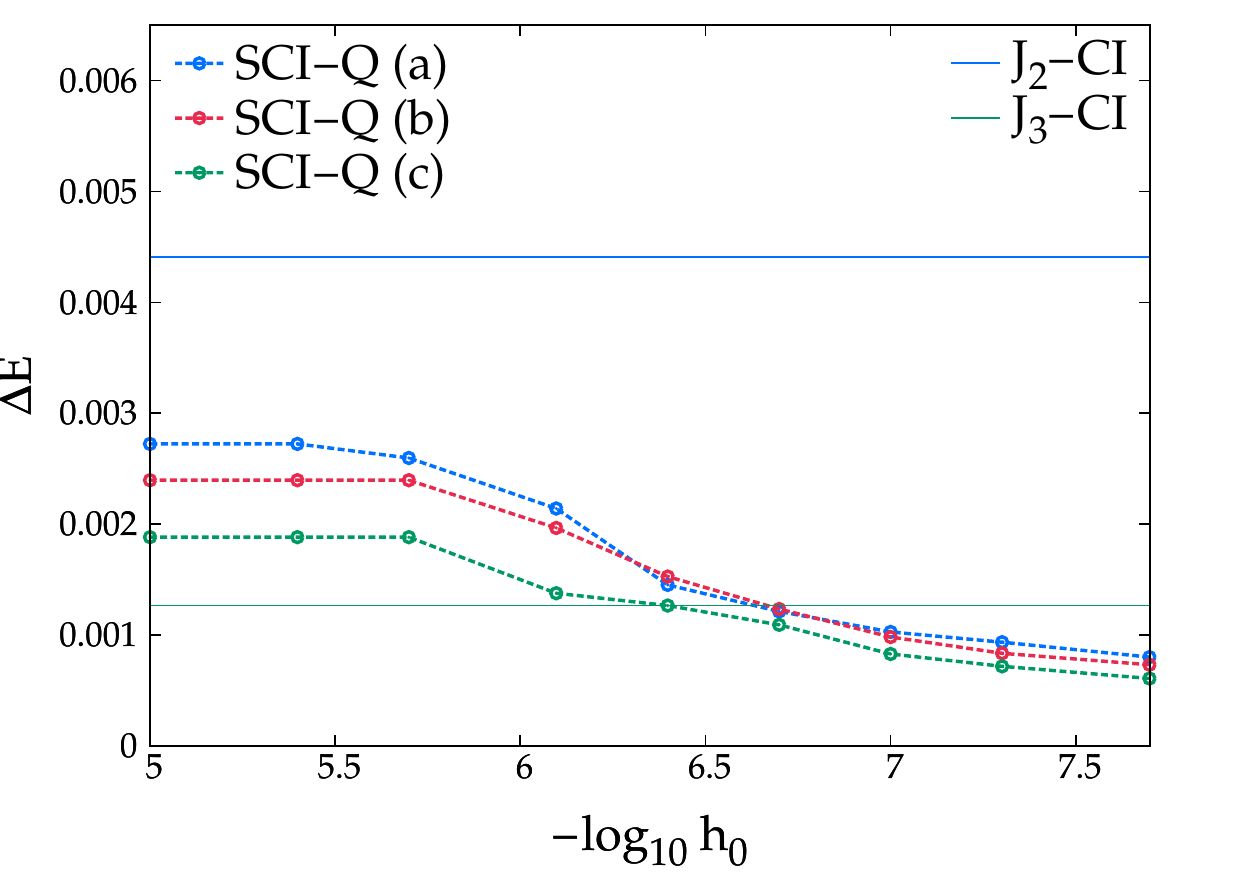}
\hfill
\includegraphics[width=\columnwidth]{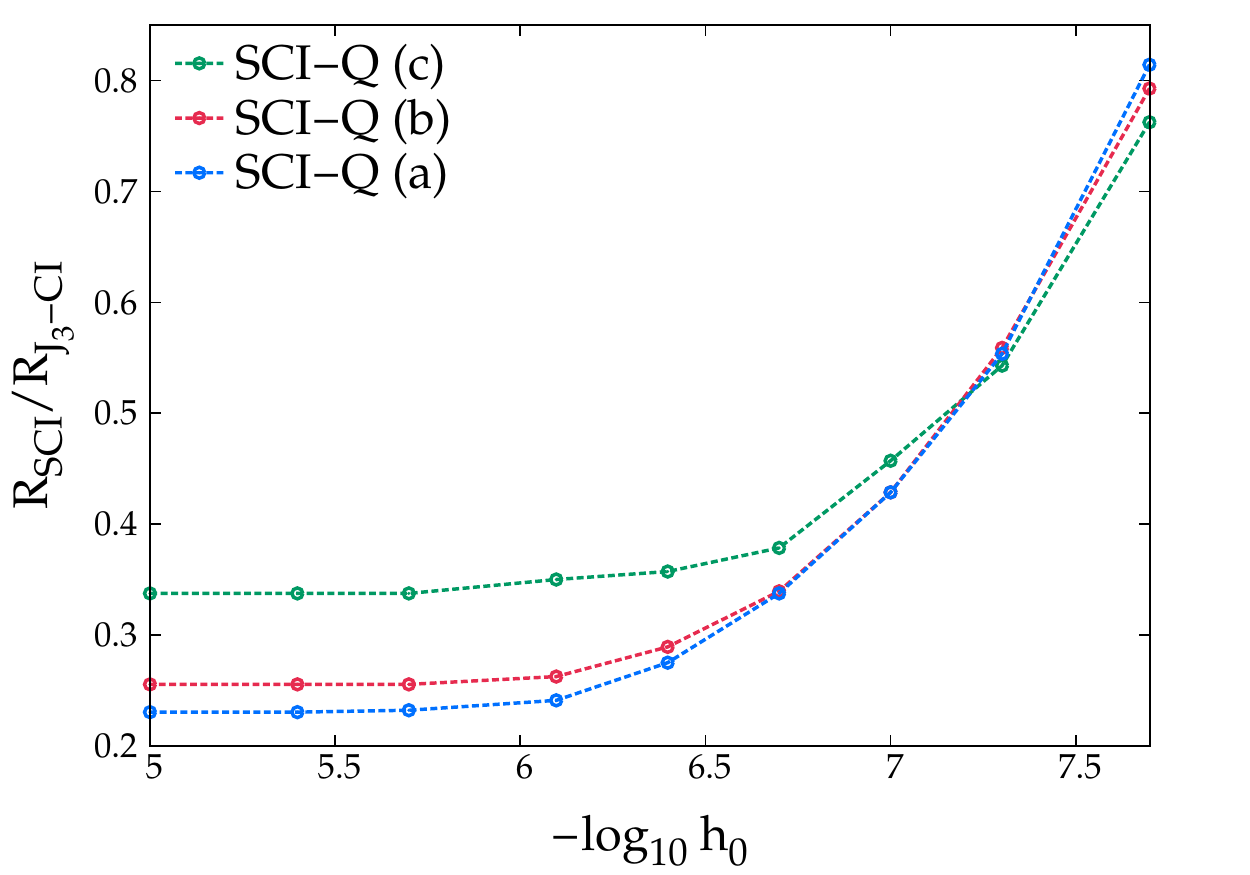}
\caption{
SCI results for the half-filled 16-level reduced BCS Hamiltonian with 
$G = 0.60$ ($G_C \sim 0.2866$). Here the Hamiltonian thresholds for different candidate elementary manifolds are chosen different. The metric threshold and the Hamiltonian threshold for elementary doubles manifold are constant for all the methods; they are $10^{-4}$ and $10^{-12}$, respectively. 
The labels `a', `b', and `c' of the SCI-Q methods indicate different Hamiltonian thresholds chosen for the elementary triples: $10^{-6}$, $5 \times 10^{-7}$, and $10^{-7}$, respectively. The 
$x$-axes for both the panels are functions of Hamiltonian thresholds for elementary quadruples, $h_{0}$.
In the $y$-axis of the left panel, we plot total energy errors 
$(E_{method} - E_{exact})$ for different SCI-Q methods. 
In the $y$-axis of the right panel, we plot total number of AGPs in the model space of different SCI-Q methods divided by the total number of states ($= 560$) in composite triples manifold, or equivalently, in $J_3$-CI.
\label{fig: en16b}}
\end{figure*}

Although LC-AGP wave functions with different pivots are variationally equivalent, the constructed spaces may have different properties. 
For example, in figure~\ref{fig: met}, we compare the metric matrices resulting from sign-flip and zero-pivot manifolds, and clearly, they have different metric densities. In addition to the unique relations between the elementary manifolds discussed in Section~\ref{subsec: fanp}, the zero-pivot metric is moderately sparse for systems that are not too strongly interacting. Using zero pivots is the simplest way to ensure that the sparsity of the metric increases for higher-order manifolds, as shown in table~\ref{tab: met} and figure~\ref{fig: met}.
The Hamiltonian matrices follow similar patterns to the corresponding metrics, and one example is shown in figure~\ref{fig: ham} for the zero-pivot case.

We use these to our advantage. Using zero-pivot manifolds, we formulate a SCI algorithm which selects energetically important states from a given set of AGPs. The procedure is outlined below:
\begin{itemize}

\item partition the AGPs into \textit{model} and \textit{candidate} spaces,

\item do LC-AGP with the model space states,

\item decide if a state from the candidate space should be chosen,

\item update model space if a new state is chosen,

\item do LC-AGP with the new model space states,

\item continue until all the trial states from the candidate space are tested.

\end{itemize}

The SCI method for LC-AGP is different from determinant-based SCI in certain important ways. The constituent AGPs are non-orthogonal. In particular, the AGPs from higher elementary manifolds are related to those from the lower manifolds, a consequence of the AGP partition introduced in eq.~\eqref{eq: part}. By a dimensionality argument, AGPs from more than two neighboring elementary manifolds are linearly dependent, so the selection procedure must make sure that some of them are excluded.
In order to guarantee linear independence  numerically, we do not include a normalized trial state $|\chi\rangle$ from the model space if the
norm of its projection $Q \: |\chi\rangle$ off the model space is smaller than a metric threshold, $m_0$. More precisely, we avoid selection of $|\chi\rangle$ from the model space if
\begin{equation}
\langle \chi| \: Q \: |\chi \rangle 
< m_0.
\end{equation}
If $Q \: |\chi\rangle$ is linearly independent to the model space, we add it to the model space if it lowers the energy significantly at lead order, 
using a Hamiltonian threshold, $h_0$. For more details, see Appendix~\ref{app: sci}.
Using smaller $m_0$ and $h_0$ values, i.e., tighter thresholds, we expect the SCI procedure to select more states while ensuring better accuracy.

We have chosen the initial model space to be composite singles so that the SCI energies remain at least as accurate as the reference AGP energies. 
Depending on the candidate space, as illustrated in figure~\ref{fig: sci}, the SCI methods are called SCI-D, SCI-T, and SCI-Q, respectively. 
We choose to test candidate AGPs beginning from the lower-order elementary manifolds. As a result, for given $m_0$ and $h_0$, SCI-D is contained in SCI-T, which in turn is contained in SCI-Q. 

Choosing a suitable metric threshold is important for ensuring linear independence in the model space while not excluding potentially important trial states. Test calculations shown in figure~\ref{fig: en12b} suggest that, given a tight Hamiltonian threshold of $10^{-12}$, SCI with increasingly tighter metric threshold rapidly approaches the corresponding $J_k$-CI energy and eventually reproduces the latter at a moderate $m_0$ value of \textit{ca.}~$10^{-4}$. Similar results hold for different regimes of correlation at $m_0 = 10^{-4}$, with the number of states in the model space close to that of the corresponding $J_k$-CI space for intermediate to large $|G|$ values, and smaller for small $|G|$ due to near-zero metric modes. \cite{GRAGP2020}
We therefore consider $m_0 = 10^{-4}$ as a good metric threshold and use it for the rest
of test calculations.

Proper choice of the Hamiltonian threshold ensures a significant portion of correlation energy
is retained by a fraction of trial states in the candidate space. As shown in Figure~\ref{fig: en16a}, lower-order elementary manifolds are energetically more important. Thus, it is sensible to choose different Hamiltonian thresholds
for different candidate elementary manifolds.
In figure~\ref{fig: en16b}, we choose the tighest Hamiltonian threshold for elementary doubles
so that the SCI energies are at least as accurate as $J_2$-CI, and less tight Hamiltonian
thresholds are used for the elementary triples and quadruples.
It is clear from figure~\ref{fig: en16b} that lower energies than $J_3$-CI can be achieved
with half of the number of states in $J_3$-CI or less upon choosing suitable Hamiltonian
thresholds.


\section{Conclusion} \label{sec: last}

We have explored expanding the many-electron wave function in a basis of non-orthogonal AGPs. We have chosen to avoid, as a first step, the task of optimizing each AGP state by generating linearly independent non-orthogonal AGP states from a given reference AGP, making LC-AGP optimization a linear problem. We have shown how different GR models (eq.~\eqref{eq: grci}) are variationally equivalent to a linear combination of AGPs that differ by one or more geminal coefficients and can also be constructed using shifted number operators acting on an AGP. 

LC-AGP provides more freedom to tune the Hilbert space than the geminal replacement models, as shown by the variational invariances of LC-AGP due to different pivots (or shifts). Indeed, the basis states of the $J_k$-CI expansion of 
eq.~\eqref{eq: jci}, equivalent to the $k$-GR model, are a limiting case of AGPs, which we have discussed in Section~\ref{subsec: fanp}. Clearly, one way to go beyond the composite manifolds is to change the geminal coefficients that do not take part in pivots while maintaining linear independence. 

We have shown how to remove energetically less important AGPs while maintaining linear independence by introducing an SCI method for non-orthogonal states, thus solving a lower-dimensional generalized eigenvalue problem while retaining a respectable portion of correlation energy. An important question is if better AGP constructions for SCI than the zero-pivot exist so that the LC-AGP wave function can be represented as compactly as possible.
Along with our past work, the benchmark results here show that the geminal based methods capture major portion of the correlation energies for the reduced BCS Hamiltonian, even with a moderate number of states. 
Extension of the present framework to include Hamiltonians that do not preserve seniority is under development and will be presented in due time. 

The use of non-orthogonal Slater determinants is well known in quantum chemistry; they are often more representative of the electronic structure and are related to symmetry broken and restored states. However, non-orthogonal determinants have the corresponding disadvantage that it is not easy to provide a mechanism by which they can systematically span the relevant Hilbert space, a task for which orthogonal Slater determinants constructed by particle-hole excitations are well suited.
The same ideas hold in the world of AGP. One can write very general and complicated geminal product models as linear combinations of non-orthogonal AGPs, but it is difficult to optimize them. If, instead, one chooses to write a basis of non-orthogonal AGPs, it may seem a priori not easy to do so in a way that spans Hilbert space. Nevertheless, this is the goal that we have here accomplished.

In this work, we have shown how we can approach various highly accurate and conceptually appealing geminal product-based CI methods via LC-AGP. 
The composite manifolds introduced here for doing LC-AGP with pivots 
systematically allow us to construct a linearly independent albeit 
non-orthogonal basis, thereby providing a rigorous alternative to the particle-hole framework.


\begin{acknowledgments}
This work was supported by the U.S. Department of Energy,
Office of Basic Energy Sciences, Computational and Theoretical Chemistry Program under Award No. DE-FG02-09ER16053. G.E.S. acknowledges support as a Welch Foundation Chair (Grant No. C-0036). We thank the reviewers for helpful comments.
\end{acknowledgments}

\appendix

\section{AGP transition RDMs} \label{app: tdm}

In this section, we will discuss computational details of evaluating AGP transition RDMs. We note in passing that expressions for transition density matrices of Richardson-Gaudin states have recently been developed by Johnson and co-workers. \cite{RGTDM2020} Since
\begin{equation}
N_p = 2 \: P_p^\dagger P_p
\end{equation}
for any seniority-zero state, \cite{TomAGPRPA2020} the transition RDMs of 
eq.~\eqref{eq: tdm} are related:
\begin{equation} \label{eq: tdm1}
Z_{\mu \nu, p}^{1, 1}
= 2 \: Z_{\mu \nu, pp}^{0, 2}.
\end{equation}
There are different ways to compute the AGP overlaps and transition RDMs.
For simplicity, we assume the geminal coefficients to be real-valued.

\subsection{ESPs and reconstruction formulae} \label{app: rf}

The overlap between AGPs $|\mu\rangle$ and $|\nu\rangle$ is an $m$-variable, 
$n$-degree ESP of $X_p^{\mu \nu} = \eta_p^\mu \: \eta_p^\nu$
\begin{subequations} \label{eq: ov1}
\begin{align}
M_{\mu \nu}
&= \sum_{1 \leq p_1 < \cdots < p_n \leq m} 
X_{p_1}^{\mu \nu} \: \cdots \: X_{p_n}^{\mu \nu}
\\
&= S_n^m \: ( \{ X_{p_i}^{\mu \nu} \: | \: 1 \leq i \leq n\} ) .
\end{align}
\end{subequations}
The elements of $\mathbf{Z}_{\mu \nu}^{0, 2}$ are related to 
$(m-2)$-variable, $(n-1)$-degree ESPs of $X_p^{\mu \nu}$
\begin{subequations} \label{eq: tdm2}
\begin{align}
&Z_{\mu \nu, \: pq}^{0, 2}
\nonumber
\\
&= \eta_p^\mu \: \eta_q^\nu
\sum_{1 \leq p_1 < \cdots < p_{n-1} \leq m-2} \:
\prod_{\substack{i = 1 \\ p_i \neq p, q}}^{n-1}
X_{p_i}^{\mu \nu}
\\
&= \eta_p^\mu \: \eta_q^\nu \: 
S_{n-1}^{m-2} \: ( \{ X_{p_i}^{\mu \nu} \: | \: 1 \leq i < n, \: p_i \neq p, q \} ).
\end{align}
\end{subequations}
Eq.~\eqref{eq: ov1} and~\eqref{eq: tdm2} can be computed by the \textit{sumESP} algorithm, as discussed in Ref.~\onlinecite{Khamoshi2019}.
Alternatively, we first compute $\mathbf{Z}_{\mu \nu}^{1, 1}$ via 
eq.~\eqref{eq: tdm2}, using eq.~\eqref{eq: tdm1}.
Then the $\mathbf{Z}_{\mu \nu}^{0, 2}$ elements can be computed using $\mathbf{Z}_{\mu \nu}^{1, 1}$. This is due to a modified version of the reconstruction formulae for AGP RDMs, \cite{Khamoshi2019} which allow expressing higher-order RDM elements in terms of lower-order ones.
We discuss it briefly below.

We assume that all the indices of a transition RDM element are different, and other cases can be transformed to this \textit{irreducible} form easily. \cite{Khamoshi2019}
Any transition RDM element with a string of number operators can be written as
\begin{align} \label{eq: rf1}
&\langle \mu| \: N_{p_1} \cdots N_{p_k} \: | \nu\rangle
\\
&= 2^{k-1} \: \sum_{i=1}^{n} \Big( \prod_{\substack{j=1 \\ j \neq i}}^{k} 
\Lambda_{p_j p_i}^{\mu \nu} \Big) \: Z_{\mu \nu, p_i}^{1, 1},
\nonumber
\end{align}
where
\begin{equation}
\Lambda_{pq}^{\mu \nu} 
= \frac{\eta_p^\mu \: \eta_q^\nu}{X_q^{\mu \nu} - X_p^{\mu \nu}}.
\end{equation}
The operator 
\begin{align} \label{eq: killer}
K_{\mu, pq}
&= (\eta_p^\mu)^2 \: P_p^\dagger \: P_q 
+ (\eta_q^\mu)^2 \: P_q^\dagger \: P_p
\\
&+ \frac{1}{2} \: \eta_p^\mu \eta_q^\mu \: (N_p N_q - N_p - N_q),
\nonumber
\end{align}
annihilates the AGP $|\mu \rangle$ \cite{TomAGPCI2019}
\begin{equation}
K_{\mu, pq} \: |\mu \rangle
= 0.
\end{equation}
If $K_{\mu, pq}^\dagger$ is the adjoint of the operator defined in 
eq.~\eqref{eq: killer}, then by solving the coupled equations below
\begin{subequations}
\begin{align}
\langle \mu| \: K_{\nu, pq} \: |\nu \rangle 
&= 0,
\\
\langle \mu| \: K_{\mu, pq}^\dagger \: |\nu \rangle 
&= 0,
\end{align}
\end{subequations}
we arrive at
\begin{align} \label{eq: rf2}
\langle \mu| \: \cdots \: P_p^\dagger P_q \: \cdots \: |\nu\rangle
&= \frac{1}{2} \: 
\frac{\eta_p^\mu \: \eta_q^\nu}{( X_p^{\mu \nu} + X_q^{\mu \nu} )}
\\
&\times \langle \mu| \: \cdots \: (N_p + N_q - N_p N_q) \: \cdots \: |\nu\rangle.
\nonumber
\end{align}
Here, we have used the facts that all the other operators in the string of 
eq.~\eqref{eq: rf2} have different indices and number operators are Hermitian.

Applying eqs.~\eqref{eq: rf1} and~\eqref{eq: rf2} to 
$\langle \mu| \: P_p^\dagger P_q \: |\nu\rangle$, we get
\begin{subequations}
\begin{align}
Z_{\mu \nu, pq}^{0, 2}
&= \frac{1}{2} \: \frac{\eta_p^\mu \: \eta_q^\nu}{X_p^{\mu \nu} + X_q^{\mu \nu}} \\
&\times \big( Z_{\mu \nu, p}^{1, 1} + Z_{\mu \nu, q}^{1, 1} 
- Z_{\mu \nu, pq}^{2, 2} \big), 
\nonumber
\\
Z_{\mu \nu, pq}^{2, 2}
&= 2 \: \big( \Lambda_{qp}^{\mu \nu} \: Z_{p, \mu \nu}^{1, 1} 
+ \Lambda_{pq}^{\mu \nu} \: Z_{q, \mu \nu}^{1, 1} \big),
\end{align}
\end{subequations}
where $Z_{\mu \nu, pq}^{2, 2} = \langle \mu| \: N_p N_q \: | \nu\rangle$.
Other AGP transition RDMs can be similarly derived using eqs.~\eqref{eq: rf1} and~\eqref{eq: rf2}.

We can thus skip computing the full transition RDM $\mathbf{Z}_{\mu \nu}^{0, 2}$  and only need $\mathbf{Z}_{\mu \nu}^{1, 1}$ and $\mathbf{\Lambda}^{\mu \nu}$ with the geminal coefficients to compute the \textbf{H} elements. Note that the reconstruction formulae are only applicable when
$X_p^{\mu \nu} \neq X_q^{\mu \nu}$.

\subsection{Transition RDMs from PBCS} \label{app: pbcsrdm}

AGP transition RDMs may be evaluated using transition RDMs of BCS states
since an $n$-pair AGP $|\mu\rangle$ can be represented as a PBCS:
\begin{equation}
|\mu\rangle 
= \mathcal{P}_{2n} \: | \Phi^\mu \rangle,
\end{equation}
where
\begin{equation}
\mathcal{P}_{2n} 
= \frac{1}{2\pi} \int_0^{2\pi} \: d\phi \: e^{i \phi (N - 2n)}
\end{equation}
is the number-projection operator to the Hilbert space of $2n$ electrons, and
\begin{equation}
|\Phi^\mu \rangle 
= \prod_p \: (u^\mu_p + v^\mu_p \: c^\dag_p c^\dag_{\bar{p}}) \: |-\rangle
\end{equation}
is a BCS state with broken number symmetry. 
Here, $u^\mu_p,\, v^\mu_p \in \mathbb{R}$
satisfy 
\begin{equation}
(u^\mu_p)^2 + (v^\mu_p)^2 = 1,
\end{equation}
and are related to $\eta^\mu_p$ by
\begin{equation}
\eta^\mu_p = \frac{v^\mu_p}{u^\mu_p}.
\end{equation}

A BCS state rotated by a gauge angle of $\phi$ is 
\begin{equation}
|\Phi^\mu(\phi) \rangle 
= e^{i\phi N} \: |\Phi^\mu \rangle.
\end{equation}
Using the idempotency of $\mathcal{P}_{2n}$, it is readily shown that\cite{Sheikh2000,PQT2011}
\begin{equation}
\langle \mu | \nu \rangle
= \frac{1}{2 \pi} \int_0^{2\pi} d\phi \: e^{-i\phi(2n)} 
\langle \Phi^\mu | \Phi^\nu (\phi) \rangle,
\end{equation}
and
\begin{align} \label{eq:z02pbcs}
&Z_{\mu \nu, pq}^{0, 2}
\\
&= \frac{1}{2 \pi} \int_0^{2\pi} d\phi \: e^{-i\phi(2n)} \: 
\langle \Phi^\mu | \Phi^\nu(\phi) \rangle \: Z^{0,2}_{\mu \nu, pq}(\phi),
\nonumber
\end{align} 
where
\begin{equation}
Z_{\mu \nu, pq}^{0, 2} (\phi)
= \frac{\langle \Phi^\mu | \: c^\dag_p c^\dag_{\bar{p}} c_{\bar{q}} c_q \: 
| \Phi^\nu (\phi) \rangle}{\langle \Phi^\mu | \Phi^\nu (\phi) \rangle}.
\end{equation}
Namely, the AGP overlap $\langle \mu | \nu \rangle$ and transition RDM 
$Z_{\mu \nu, pq}^{0, 2}$ are expressed in terms of the BCS overlap 
$\langle \Phi^\mu | \Phi^\nu (\phi) \rangle$ and transition RDM $Z^{0,2}_{\mu\nu, pq}(\phi)$, respectively. 

The BCS overlap is a pfaffian,\cite{Robledo2009} which in this case can be further simplified into
\begin{equation}
\langle \Phi^\mu | \Phi^\nu (\phi) \rangle 
= \prod_p \: e^{i\phi} \: \sigma^{\mu\nu}_p (\phi),
\end{equation}
where
\begin{equation}
\sigma_p^{\mu\nu} (\phi) 
= e^{-i\phi} \: (u_p^\mu)^2 + e^{i\phi} \: (v_p^\mu)^2.
\end{equation}
The BCS transition RDM $Z_{\mu \nu, \: pq}^{0, 2} (\phi)$ may be decomposed by 
generalized Wick's theorem. \cite{GenWicks} 
This leads to
\begin{equation} \label{eq:z02bcs}
Z_{\mu \nu, pq}^{0, 2} (\phi)
= \delta_{p q} \: \big( \rho^{\mu\nu}_p(\phi) \big)^2
+ \bar{\kappa}^{\mu\nu*}_p (\phi) \: \kappa^{\mu\nu}_q (\phi),
\end{equation}
where
\begin{subequations}
\begin{align}
\rho_p^{\mu\nu} (\phi)
&= \frac{\langle \Phi^\mu | \: c^\dag_p c_p \: |\Phi^\nu (\phi) \rangle}{\langle \Phi^\mu | \Phi^\nu (\phi) \rangle}
\\
&= e^{i\phi} \: v^{\nu}_p \: \sigma^{\mu\nu}_p (\phi)^{-1} \: v^\mu_p,
\\
\kappa_p^{\mu\nu} (\phi)
&= \frac{\langle \Phi^\mu | \: c_{\bar{p}} c_p \: |\Phi^\nu (\phi) \rangle}{\langle \Phi^\mu | \Phi^\nu (\phi) \rangle}
\\
&= e^{i\phi} \: v^{\nu}_p \: \sigma^{\mu\nu}_p (\phi)^{-1} \: u^\mu_p,
\\
\bar{\kappa}_p^{\mu\nu*} (\phi)
&= \frac{\langle \Phi^\mu | \: c^\dag_p c^\dag_{\bar{p}} \: |\Phi^\nu (\phi) \rangle}{\langle \Phi^\mu | \Phi^\nu (\phi) \rangle}
\\
&= e^{-i\phi} \: u^{\nu}_p \: \sigma^{\mu\nu}_p (\phi)^{-1} \: v^\mu_p
\end{align}
\end{subequations}
are one-body BCS transition RDMs, which have been represented as diagonal matrices. These quantities are of $m \times l$ dimension, where $l$ denotes the size of the numerical quadrature for the gauge integration. 
Therefore, eqs.~\eqref{eq:z02pbcs} and~\eqref{eq:z02bcs} provide a decomposed expression of $\mathbf{Z}_{\mu \nu}^{0,2}$ that facilitates robust and efficient computation. Expressions of other AGP transition RDMs can be derived similarly.
In practice, we do not need to construct full AGP transition RDMs; instead, 
$\bm{\rho}^{\mu \nu}$, $\bm{\kappa}^{\mu\nu}$, and $\bar{\bm{\kappa}}^{\mu\nu}$ can be directly contracted with other tensors.


\section{Special cases of linear dependence} \label{app: lin}

Without freezing an level, linear dependence of states in form of 
eq.~\eqref{eq: jbar} may arise. Two special cases of such linear dependence are shown here, which are by no means exhaustive.
Here we restrict ourselves to identical shifts $\beta_p^\mu = \beta$, and define
\begin{equation}
\bar{N}_p = N_p + \beta.
\end{equation}
Eq.~\eqref{eq: jbar} then becomes
\begin{equation} \label{eq: jbarsym}
|n; \: p_1 \cdots p_k\rangle
= \frac{1}{\beta^k} \prod_{i=1}^k \: \bar{N}_{p_i} \: |n\rangle.
\end{equation}

\subsection{Totally symmetric states}

The totally symmetric linear combination of the states defined in 
eq.~\eqref{eq: jbarsym} is special because it may be generated by acting polynomials of the total number operator $N$ 
(eq.~\eqref{eq: totN}) on the reference AGP $|n\rangle$. Since $|n\rangle$ is an eigenfunction of $N$ (eq.~\eqref{eq: totNonAGP}), it follows that this totally symmetric state is a multiple of $|n\rangle$. Linear dependence arises when the coefficient is zero.

We first define the $k$-degree ESP of the shifted number operator
\begin{equation} \label{eq: espN}
S^m_k
= \sum_{1 \leq p_1 < \cdots < p_k \leq m} \bar{N}_{p_1} \cdots \bar{N}_{p_k},
\end{equation}
where $S^m_k$ is short for 
$S_k^m \: (\{ N_{p_i} \: | \: 1 \leq i \leq k \})$.
The totally symmetric state is then written as
\begin{equation}
S^m_k \: |n\rangle
= \beta^k \sum_{1 \leq p_1 < \cdots < p_k \leq m} |n; \: p_1 \cdots p_k\rangle,
\end{equation}
using eq.~\eqref{eq: jbarsym} and~\eqref{eq: espN}. 
We may readily verify that 
\begin{subequations}
\begin{align}
    S^m_0 \: |n\rangle
    &= |n\rangle, \\
    \label{eq: sm1}
    S^m_1 \: |n\rangle
    &= ( 2n + m \beta ) \: |n\rangle, \\
    \label{eq: sm2}
    S^m_2 \: |n\rangle
    &= \frac{1}{2} \: \big[
        m (m - 1) \: \beta^2 +
        4n (m - 1) \: \beta 
        \\
        &+ 4n (n - 1)
    \big] \: |n\rangle.
    \nonumber
\end{align}
\end{subequations}
From eq.~\eqref{eq: sm1}, the states in
$\{ |n; \: p\rangle \: | \: 1 \leq p \leq m \}$ are linearly dependent
when 
\begin{equation}
\beta = -\frac{2n}{m}.
\end{equation}
Similarly, using eq.~\eqref{eq: sm2}, linear dependence arises for the states in 
$\{ |n; \: pq \rangle \: | \: 1 \leq p < q \leq m \}$ when
\begin{equation}
\beta 
= \frac{- 2 \: \big[ n (m-1) \pm \sqrt{n (m-1) (m-n)} \big]}{m (m-1)}.
\end{equation} 

Generally, the following recursive relation holds:
\begin{align}  \label{eq: recursion}
&(k+1) \: S^m_{k+1} \: |n\rangle 
\\
&+ \big[ (2k - m) \: \beta + 2k - 2n \big] \: S^m_k \: |n\rangle
\nonumber
\\
&+ (k - m - 1) \: (\beta^2 + 2\beta) \: S^m_{k-1} \: |n\rangle 
= 0,
\nonumber
\end{align}
which can be used to write out the expression of $S^m_k \: |n\rangle$ for any 
$k \geq 2$. We have used eq.~\eqref{eq: idem}, or 
\begin{equation}
\bar{N}_p^2 = 2 \: (\beta + 1) \: \bar{N}_p + \beta^2 + 2\beta,
\end{equation}
for seniority-zero states, in the derivation of eq.~\eqref{eq: recursion}.

\subsection{A special case for sign-flip}

For the sign-flip manifolds ($\beta = -1$), we also observe linear dependence at $2k = m$ as
\begin{equation}
\prod_{i=1}^k \bar{N}_{p_i} \: |n\rangle
= e^{i \pi n} \prod_{i=k+1}^m \bar{N}_{p_i} \: |n\rangle.
\end{equation}
This relation can be shown by observing
\begin{equation}
(-1)^k \prod_{i=1}^k \bar{N}_{p_i} \: |n\rangle
= e^{\pm i \frac{\pi}{2} \sum_{i=1}^k N_{p_i}} \: |n\rangle,
\end{equation}
which is a special case of eq.~\eqref{eq: ej1}.


\section{Non-orthogonal SCI criteria} \label{app: sci}

Suppose we have a collection of non-orthogonal states $\{ |p\rangle, |q\rangle, \cdots \}$, which we call the model space. We want to decide if a non-orthogonal state $|\mu\rangle$ should be added to the model space. We assume all the states are normalized.

First, we want to know if $|\mu\rangle$ is linearly independent of the model space. This can be accomplished by using the projector
\begin{equation}
Q = I - \sum_{pq} \: |p\rangle \: 
X_{pq} \: \langle q|,
\end{equation}
where \textbf{X} is inverse of the model space metric
\begin{equation}
\mathbf{X} = \mathbf{S}^{-1}.
\end{equation}
The metric of the expanded space, with the projection of $|\mu \rangle$ off the model space included is
\begin{equation}
\mathbf{M} = 
\begin{pmatrix}
\mathbf{S} & \mathbf{0} \\
\mathbf{0} & \langle \mu | \: Q \: | \mu\rangle 
\end{pmatrix}.
\end{equation}
We have our first criterion that the state must pass:
\begin{equation} \label{eq: mtest}
\langle \mu| \: Q \: | \mu \rangle
> m_0,
\end{equation}
where $m_0$ is a small real number called the metric threshold.
This guarantees that $Q \: |\mu\rangle$ is normalizable with good precision, and hence $|\mu\rangle$ is numerically linearly independent.

If $|\mu\rangle$ passes the metric test (eq.~\eqref{eq: mtest}), that means it could be added to the model space. However, we need to decide if we should add it, based on a Hamiltonian test that we discuss now. 
Let us assume that the Hamiltonian in the model space has normalized eigenvector
\begin{equation}
|\psi\rangle
= \sum_p \: C_p \: |p\rangle,
\end{equation}
with eigenvalue $E$. 
Then in the basis of $\{ |\psi\rangle, \: Q \: |\mu\rangle \}$, the metric and Hamiltonian matrices are
\begin{subequations}
\begin{align}
\mathbf{M} &= 
\begin{pmatrix}
1 & 0 \\
0 & \langle \mu | \: Q \: | \mu\rangle 
\end{pmatrix},
\\
\mathbf{H} &= 
\begin{pmatrix}
E & \langle \psi| \: H \: Q \: |\mu\rangle \\
\langle \mu| \: Q \: H \: |\psi\rangle & 
\langle \mu | \: Q \: H \: Q \: | \mu\rangle
\end{pmatrix}.
\end{align}
\end{subequations}
For brevity, we define
\begin{subequations}
\begin{align}
\bar{M} &=
\langle \mu | \: Q \: | \mu\rangle,
\\
\bar{T} &= 
\langle \psi | \: H \: Q \: | \mu\rangle,  
\\
\bar{H} &= 
\langle \mu | \: Q \: H \: Q \: | \mu\rangle.
\end{align}
\end{subequations}
Then the smallest generalized eigenvalue of 
$ \{ \mathbf{H}, \: \mathbf{M}\} $ is
\begin{subequations}
\begin{align}
\varepsilon
&= \frac{1}{2 \: \bar{M}} \: 
\big( \bar{H} + E \: \bar{M} - \bar{R} \big),
\\
\bar{R}
&= \sqrt{ (\bar{H} - E \: \bar{M})^2  + 4 \: \bar{M} \: \bar{T}^2 }. 
\end{align}
\end{subequations}
The new state should be added to the model space if the energy correction is sufficiently large. Hence to pass the Hamiltonian test, the following must be true
\begin{equation}
\Big\vert \frac{\varepsilon - E}{E} \Big\vert \: > \: h_0,
\end{equation}
where $h_0$ is a small real number called the Hamiltonian threshold. 


\section*{DATA AVAILABILITY}

The data that support the findings of this study are available from the corresponding author upon reasonable request.

\bibliography{LCAGP}

\end{document}